\documentclass[english,english]{emulateapj}
\usepackage[T1]{fontenc}
\usepackage[latin1]{inputenc}
\usepackage{graphicx}
\usepackage{amssymb}

\makeatletter

\usepackage{graphicx}
\usepackage{amssymb}

\usepackage{times}
\usepackage{amsmath}

\shorttitle{PROBING POST-NEWTONIAN GRAVITY}
\shortauthors{ZUCKER ET AL.}

\usepackage{babel}

\usepackage{babel}
\makeatother
\begin{document}

\title{Probing Post-Newtonian Gravity near the Galactic Black Hole\\
 with Stellar Doppler Measurements}

\author{Shay Zucker\altaffilmark{1}, Tal Alexander\altaffilmark{1,2}, Stefan
Gillessen\altaffilmark{3}, Frank Eisenhauer\altaffilmark{3} and
Reinhard Genzel\altaffilmark{3} }

\email{shay.zucker@weizmann.ac.il, tal.alexander@weizmann.ac.il, 
ste@mpe.mpg.de, eisenhau@mpe.mpg.de,\\ genzel@mpe.mpg.de}

\altaffiltext{1}{Faculty of Physics, Weizmann Institute of Science, PO
Box 26, Rehovot 76100, Israel}
\altaffiltext{2}{The William Z. \& Eda Bess Novick career development
chair}
\altaffiltext{3}{Max-Planck-Institut f\"{u}r extraterrestriche Physik,
Postfach 1312, Garching D-85741, Germany}

\begin{abstract}
Stars closely approaching the massive black hole in the center of
the Galaxy provide a unique opportunity to probe post-Newtonian physics
in a yet unexplored regime of celestial mechanics. Recent advances
in infrared stellar spectroscopy allow the precise measurement of
stellar Doppler shift curves and thereby the detection of $\beta^{2}$
post-Newtonian effects (gravitational redshift in the black hole's
potential and the transverse Doppler shift). We formulate a detection
procedure in terms of a simplified post-Newtonian parametrization.
We then use simulations to show that these effects can be decisively
detected with existing instruments after about a decade of observations.
We find that neglecting these effects can lead to statistically significant
systematic errors in the derived black hole mass and distance. 
\end{abstract}


\section{Introduction}

General Relativity (GR) is the least tested theory of the four fundamental
forces of nature. The $m\!\sim\!3$--$4\!\times\!10^{6}\, M_{\odot}$
massive black hole (MBH) in the Galactic center (GC) and the stars
around it (\citealt{Eisetal2005}; \citealt{Gheetal2005}) provide
a unique laboratory for detecting post-Newtonian (PN) effects and
for probing GR (for a review, see \citealt{Ale2005}). Long-term astrometric
and spectroscopic monitoring of stars orbiting the MBH allow the derivation
of their orbital elements. The stars are observed at periapse distances
as small as $r_{p}\!\lesssim\!\!10^{3}r_{s}$ ($r_{s}\!\equiv\!2Gm/c^{2}$)
with velocities as high as $\beta_{p}\!\sim\!\mathrm{few}\!\times\!0.01$
($\beta\!\equiv\! v/c$). The relativistic parameter at periapse,
$\Upsilon\!\equiv\! r_{s}/r_{p}\!\gtrsim\!10^{-3}$ is a few times
higher than that on the surface of a white dwarf. This is an unexplored
regime of celestial mechanics. For comparison, the high precision
confirmation of GR predictions in the Hulse-Taylor binary pulsar (PSR
1913+16) was for masses of $\sim\!1.4\, M_{\odot}$ with $\beta_{p}\!\sim\!0.003$
at $r_{p}\!\sim\!2\!\times\!10^{5}r_{s}$ (\citealt{TayWei1989}).
Note that although some accretion processes occur at $\Upsilon\!\gtrsim\!{\cal {O}}(0.1)$,
the complex physics of accretion and its many uncertainties severely
limit the usefulness of accretion emission as a probe of GR.

At present all the observed orbits can be adequately modeled in terms
of Newtonian physics. With improved resolution, higher precision and
a longer baseline, deviations from Keplerian orbits may be detectable.
In this \emph{Letter} we examine the future detectability of leading
order ($\beta^{2}$) PN effects with \emph{existing} instrumental
capabilities. We show that such effects can be detected in the observed
stellar Doppler shift curves (note that the Doppler shift is no longer
equivalent to the radial velocity when PN effects are included).

The measurement of stellar proper motions and accelerations in the
inner arcsecond ($\mathrm{0.04\, pc}$) of the GC (\citealt{EckGen1996};
\citealt{Gheetal1998}) preceded by more than a decade the measurement
of radial velocities for these stars (\citealt{Gheetal2003}; \citealt{Eisetal2003a}).
Thus, most theoretical studies of the detectability of PN orbital
effects focused on those that affect the proper motions, such as the
periapse shift (e.g. \citealt{RubEck2001}; \citealt{Weietal2005}).
However, with the advent of adaptive-optics-assisted IR imaging spectroscopy
(e.g. VLT/SINFONI; \citealt{Eisetal2003b}), it is now the observed
Doppler shifts rather than the proper motions that provide the tightest
constraints on deviations from Newtonian orbits. The current quality
of stellar spectroscopy in the GC allows the determination of the
Doppler shift, $z$, to within $\delta z\!\sim\!25\,\mathrm{km\, s^{-1}}/c$,
or $\delta\lambda/\lambda\!\sim\!10^{-4}$ (\citealt{Eisetal2005}).

The observed Doppler shift curve $z(t)$ can be expanded in terms of
the 3D source velocity $\beta(t^{\prime})$, where $t^{\prime}$ and $t$
are the light emission and arrival times, respectively,
\begin{equation}
z=\Delta\lambda/\lambda=B_{0}+B_{1}\beta+B_{2}\beta^{2}+{\cal
{O}}(\beta^{3})\,.\label{e:z}\end{equation} As shown in
\S\ref{s:effects}, two effects contribute to the 2nd-order term
(\citealt{KopOze1999}), $B_{2}\!=\! B_{2,t}\!+\! B_{2,g}$: the special
relativistic transverse Doppler effect, with $B_{2,t}\!=\!1/2$, and
the GR gravitational redshift, with $B_{2,g}\!=\!1/2$. It then follows
that the 2nd-order PN effects are detectable with existing
instruments, since $(B_{2,t}\!+\! 
B_{2,g})\beta_{p}^{2}\!\sim\!10^{-3}\!>\!\delta\lambda/\lambda\!\sim\!10^{-4}$.

This \emph{Letter} is organized as follows. In \S\ref{s:effects}
we discuss the properties of the $\beta^{2}$ terms and the implications
for their detectability. In \S\ref{s:simulations} we demonstrate
the detectability of these effects by simulated observations.  We
discuss our results in \S\ref{s:conclusion}.

\section{$\beta^{2}$ effects in the Doppler shift}

\label{s:effects}

When $\Upsilon\!\ll\!1$ the orbit is Keplerian to a good approximation
and the orbital energy equation (the vis viva equation) is \begin{equation}
\beta^{2}\simeq r_{s}/r-r_{s}/2a\,,\label{e:b2}\end{equation}
 where $a$ is the orbital semi-major axis. For an eccentric orbit
$r_{p}\!\ll\! a$ and so near periapse $\Upsilon\!\sim\!\beta^{2}$.
Two effects contribute to the $\beta^{2}$ term of the Doppler shift
expansion (Eq. \ref{e:z}). One is the GR gravitational redshift,
\begin{equation}
z_{g}\equiv\left(1-r_{s}/r\right)^{-1/2}-1\simeq r_{s}/2r\,,\end{equation}
 which also contributes a constant to the 0th order term of the $z$-expansion,
$z_{g}\!=r_{s}/4a\!+\!(1/2)\beta^{2}=B_{0,g}\!+\! B_{2,g}\beta^{2}$
(Eq. \ref{e:b2}). The second is the full relativistic Doppler shift
of a moving source seen by an observer at rest,

\begin{equation}
z_{D}\equiv(1\!+\!\beta\cos\vartheta)\left/\sqrt{1\!-\!\beta^{2}}\right.-1\simeq\beta\cos\vartheta+\beta^{2}/2\,,\end{equation}
 where $\vartheta$ is the angle between the velocity vector and the
line of sight. The Newtonian Doppler shift, $z_{r}\!\equiv\!\beta\cos\vartheta\!=\! B_{1}\beta$,
is the 1st- order term in the $z$-expansion. The special-relativistic
transverse Doppler effect, $z_{t}\!\equiv\!\beta^{2}/2\!=\! B_{2,t}\beta^{2}$,
contributes to the 2nd-order term in the $z$-expansion. 

The light travel time from the star to the observer changes with the
orbital phase in an inclined orbit. We include this effect, known as
the R\o{}mer delay, by relating the observed Doppler shift $z(t)$ to
the actual velocity at emission \linebreak
$\beta\{t\!-\![R_{0}\!-\!r(t^{\prime})\cos\psi(t^{\prime})]/c\}$,
$\psi$ is the angle between $\mathbf{r}$ and the line of sight and
$R_{0}$ is the distance to the MBH. It is crucial to do so because the
phase dependence of the light travel time also contributes to the 2nd
order term (e.g. \citealt{Ale2005}).

\begin{figure}
\includegraphics[%
  width=1.0\columnwidth,
  keepaspectratio]{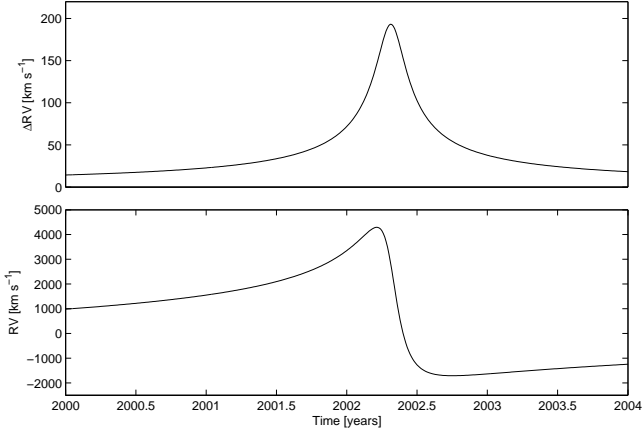}

\caption{\label{f:effects}Bottom: the full relativistic radial velocity curve
of S2 near periapse. Top: the contribution of the post-Newtonian $\beta^{2}$
effects of gravitational redshift and transverse Doppler shift to
the total. }
\end{figure}

The gravitational redshift and the transverse Doppler shift affect
only the measured Doppler shifts, but not the astrometric proper motions.
These are treated here as Newtonian, since astrometric PN effects
are below the current astrometric precision. Figure \ref{f:effects}
shows the 2nd-order PN effects that are predicted for the star S2,
based on the orbital elements derived for it from a Newtonian fit
to the data (\citealt{Eisetal2005}). As expected, the PN effects
near periapse are a few percents of the total measured Doppler shift,
since for $z_{t}\!+\! z_{g}\!\ll\!1$,\begin{equation}
(z_{t}\!+\! z_{g})/z\sim\beta_{p}^{2}/\beta_{p}\sim\beta_{p}\,.\label{e:PN}\end{equation}

\citet{Bruetal1975} studied the detectability of the gravitational
redshift and transverse Doppler shift effects in binary pulsar timing
data. They show that radial velocity data alone are not enough to
detect these effects, since to leading order they can be absorbed
in a purely Newtonian solution by very small modifications of the
orbital elements $K$ (radial velocity amplitude), $\omega$ (argument
of the periapse) and $V_{0}$ (the line-of-sight center-of-mass velocity).
To break the degeneracy they suggest using the GR periapse shift ($\dot{{\omega}}\!\neq\!0$).
While this may be feasible for a binary pulsar, where the orbital
period is very short and the timing is extremely accurate, a different
approach is required for the orbits around the Galactic MBH. In this
context there are two possibilities to break the degeneracy. One is
to use the astrometric data, which can constrain both $\omega$ and
$K$. Another is to use the fact that the stars all orbit a common
center-of-mass and share the same $V_{0}$. Thus, a combined solution
of the astrometric and radial-velocity orbits for all the stars can
break the degeneracy and reveal the 2nd-order PN effects.

Best-fit Newtonian orbital elements and their errors are usually obtained
by a $\chi^{2}$ fit of Newtonian orbits to the data. We extend this
procedure by introducing a new parameter, $X_{2}$, which quantifies
the strength of the 2nd-order PN effects, so that the GR gravitational
redshift term is parametrized as \begin{equation}
z_{g}(X_{2})=\left(1-X_{2}r_{s}/r\right)^{-1/2}-1\,,\end{equation}
 and the full relativistic Doppler term as \begin{equation}
z_{D}(X_{2})=\left(1+\beta\cos\vartheta\right)\left/\sqrt{1-X_{2}\beta^{2}}\right.-1\,.\end{equation}
 For Newtonian physics, $X_{2}\!=\!0$, whereas for PN GR, $X_{2}\!=\!1$.
Values other than zero or one are also possible in principle and would
indicate deviations from the predictions of both theories. The $X_{2}$
parametrization is chosen here for its simplicity, but is neither
general nor unique. For example, it arbitrarily presupposes that deviations
from GR (expressed by $z_{g}$) are related in a specific way to deviations
from special relativity (expressed by $z_{D}$). If future data quality
or candidate alternative theories warrant it, the fit procedure can
be easily generalized to include multiple PN parameters.

We looked for evidence of PN effects in the existing orbital data
by applying the PN fit procedure to the orbits of the 6 stars S1,
S2, S8, S12, S13 and S14 (\citealt{Eisetal2005}). We obtain a best-fit
value of $X_{2}=-4.9\pm2.9$, which appears to exclude PN GR at a
$0.02$ confidence level and Newtonian physics at a $0.05$ confidence
level. The number of currently available Doppler shift measurements
is very small, and the astrometric data is not homogeneous. It is
conceivable that these results are strongly influenced by systematic
effects. Future observations will settle this question.

\section{Simulations}

\label{s:simulations}

\begin{figure}
\includegraphics[%
  width=1.0\columnwidth,
  keepaspectratio]{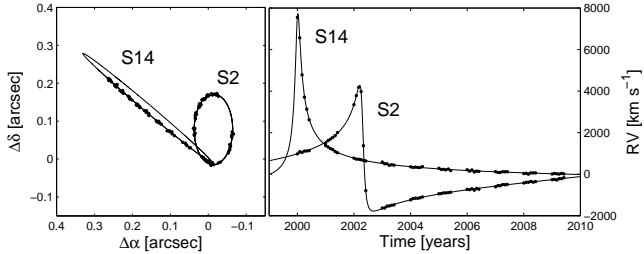}

\caption{\label{f:orbits} Simulated orbits of the stars S2 and S14. The {}``observed''
astrometric data (left) and Doppler shift data (right) are marked
by points.}
\end{figure}

We performed a suite of simulations to investigate the detectability
of 2nd-order PN effects in the Doppler shift data with existing instrumental
capabilities and with realistic astrometric and spectroscopic precision.
Since the relative strength of the PN effects scales with $\beta_{p}$
(Eq. \ref{e:PN}), we focus on stars that pass through deep (high
velocity) periapse during the monitoring period. This has actually
happened in the period 2000--2003 (\citealt{Eisetal2005}; Fig. \ref{f:orbits})
for the two stars S2 (with period $P\!=\!15.2\,\mathrm{yr}$, eccentricity
$e\!=\!0.88$, $r_{p}\!=\!1500r_{s}$, $\beta_{p}\!=\!0.02$) and
S14 (with $P\!=\!38\,\mathrm{yr}$, $e\!=\!0.94$, $r_{p}\!=1400r_{s}$,
$\beta_{p}\!=\!0.03$). Unfortunately, at that time precise stellar
spectroscopy of the inner GC was just becoming available, and only
a few Doppler measurements of S2 were obtained (\citealt{Gheetal2003};
\citealt{Eisetal2003a}).

To make the simulations realistic, we used the $6$ orbits that were
actually monitored and solved by \citet{Eisetal2005}. We generated
synthetic orbits based on the published Newtonian orbital elements.
The simulated datasets consisted of one observation per month ($2$
astrometric coordinates and one Doppler shift per star), for 6 consecutive
months per year (the GC is observable only part of the year). The
starting point was the year 2000, so the first $3$ years already
cover the periapse passage of S2 and S14 (Fig. \ref{f:orbits}). To
each measurement we added uncorrelated random Gaussian errors, with
a standard deviation of $25\,\mathrm{km}\,\mathrm{s}^{-1}$ for the
radial velocities and $1.5\,\mathrm{mas}$ per coordinate for the
astrometry, which are representative of the measurement errors for
S2. The MBH's astrometric position and proper motion were assumed
to be known, since frequent accretion flares can reveal the MBH's
location (\citealt{Gen2003b}; \citealt{Ghe2004}). These assumptions
are somewhat optimistic, since S2 is better measured than most stars,
and since accurate localization of the MBH by the flares is still
challenging. However, judging by the continuing improvement in the
data acquisition, it is not unreasonable to expect such data quality
in the near future. 

We generated progressively longer datasets, starting from $3$ years
and up to $18$ years. We repeated the entire procedure twice, both
with and without the PN effects. We then applied the PN orbital fit
procedure to the simulated datasets, and recorded the best-fit values
of $X_{2}$, as well as those of $m$ and $R_{0}$.

\begin{figure}
\begin{center}\begin{tabular}{c}
\includegraphics[%
  width=1.0\columnwidth,
  keepaspectratio]{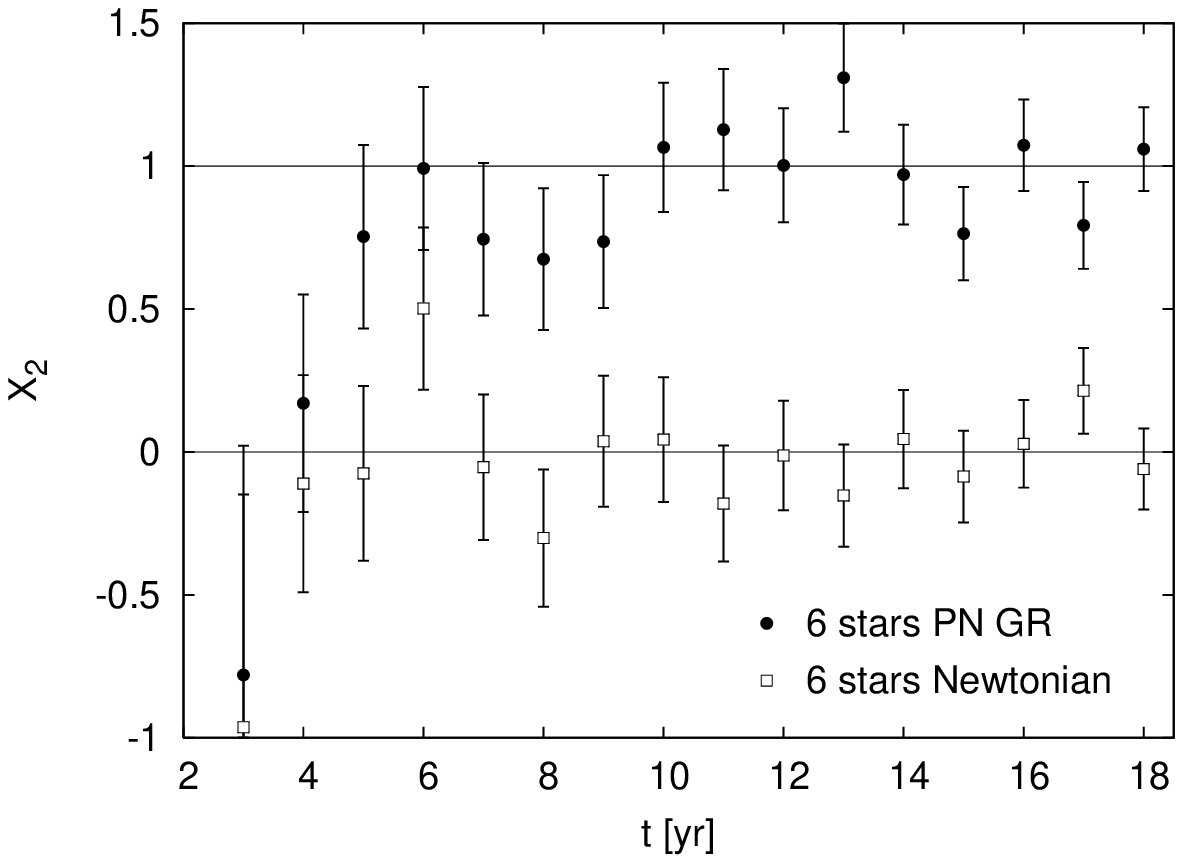}\tabularnewline
\includegraphics[%
  width=1.0\columnwidth,
  keepaspectratio]{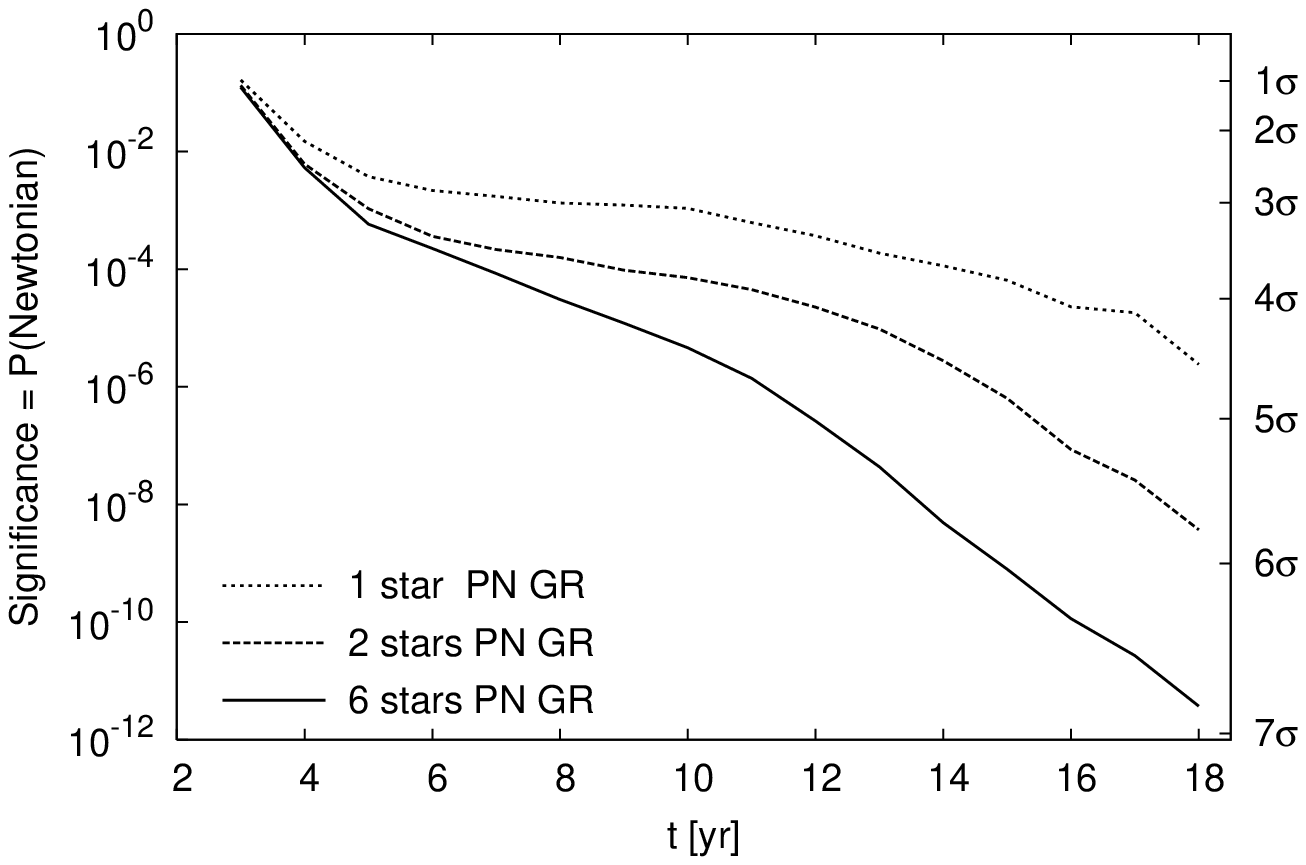} \tabularnewline
\end{tabular}\end{center}

\caption{\label{f:sim} Top: $1\sigma$ limits on the best-fit $X_{2}$ parameter
that quantifies the strength of the leading order PN effects in the
data ($X_{2}\!=\!0$ for Newtonian physics and $X_{2}\!=\!1$ for
PN physics). Bottom: The mean significance of detection of the PN
effects (rejection of Newtonian hypothesis) in the simulated dataset
with PN physics, as function of monitoring time for $1$, $2$ and
$6$ stars, presented in terms of the rejection probability (left
axis) and as standard deviations (right axis).}
\end{figure}

The upper panel of Figure \ref{f:sim} shows the derived values of
$X_{2}$ for both Newtonian and PN simulated orbits of the 6 stars
(of which only S2 and S14 go through deep periapse). The errors on
$X_{2}$ clearly improve with longer monitoring. The two hypotheses
can be decisively distinguished by the data at a $4\sigma$ confidence
level after $10$ years of monitoring. The lower panel translates
the errors on $X_{2}$ into expected detection significance for $1$
(S2 only), $2$ (S2 and S14), and all $6$ stars. These confidence
levels reflect the assumed measurement errors. The effect of a change
in the assumed errors can be easily assessed by noting that a global
scaling of all errors simply scales the confidence interval by the
same factor. 

\section{Discussion}

\label{s:conclusion}

We demonstrated that leading order PN effects in the Doppler shifts
of stars orbiting the Galactic MBH can be decisively detected in about
a decade of astrometric and spectroscopic monitoring with existing
instruments. We added to the usual set of Keplerian orbital elements
a simple PN parameter, $X_{2}$, which quantifies the strength of
the PN effects. This PN parameter can be easily generalized to a set
of PN parameters to describe any extension of Newtonian physics. 

It is important to optimize the observing strategy to improve the
chances of detecting PN effects. Generally, the observational program
should focus on deep periapse passages and include as many stellar
orbits as possible to better constrain the parameters and break the
Newtonian/PN degeneracy. Our simulations were all based on the currently
observed orbits and assumed optimal observing conditions. However,
future observations will include additional orbits and be carried
under varying observing conditions, which were not considered here.
A comprehensive study of a range of orbital configurations, measurement
errors and observational resources allocation strategies remains to
be done. 

Radial velocity information allows the unambiguous derivation of both
the MBH mass, $m$, and distance to the GC, $R_{0}$, without prior
assumptions (\citealt{Sal1999}). If ignored, the PN effects can introduce
systematic errors in the best-fit values of these parameters. For
example, the inclusion of the gravitational redshift, transverse Doppler
and R\o{}mer effects (not taken into account in previous orbital
solutions) in the orbital fitting of existing data changes the best-fit
values from $m\!=\!(3.47\pm0.25)\!\times\!10^{6}\, M_{\odot}$ to
$m\!=\!(3.65\pm0.28)\!\times\!10^{6}\, M_{\odot}$ and from $R_{0}\!=\!7.47\pm0.24\,\mathrm{kpc}$
to $R_{0}\!=\!7.59\pm0.25\,\mathrm{kpc}$. These changes are not yet
statistically significant. However, our simulations indicate that
once Doppler shift measurements are available, the discrepancies can
become highly significant. For example, the neglect of the PN effects
in the orbital solutions of the simulated data of $6$ stars monitored
for a decade results in $8\sigma$ discrepancies in the derived values
of $m$ and $R_{0}$.

Another closely related post-Newtonian effect is phase-dependent stellar
flux variability. This is due to the combined effects of relativistic
beaming and the Doppler shift. To leading order, beaming causes the
bolometric flux to scale as $1\!-\!4\beta\cos\vartheta$ (\citealt{Ryb1979},
Eq. 4.97b). However, since the infrared bands lie in the Rayleigh-Jeans
part of typical S-star spectra, the flux in the emitted spectral range
that is observed after being Doppler-shifted scales as $1\!+\!3\beta\cos\vartheta$.
Therefore the total variability scales as $1\!-\!\beta\cos\vartheta$,
which translates to an expected variability near periapse of order
$\beta_{p}\!\sim{\cal {O}}(0.01)$. 

We note that the effects of any dark distributed mass around the MBH
on the Doppler shift is likely to be small, but perhaps not completely
negligible. The change in the Doppler shift at periapse, $\delta z{}_{\mathrm{dm}}$,
due to the retrograde periapse shift, $\delta\omega_{\mathrm{dm}}$,
that is induced by the gravitational potential perturbation of the
dark mass (see \citealt{Ale2005}) is of the order $\delta z{}_{\mathrm{dm}}\!\sim\!{\cal {O}}(\left|\delta\omega_{\mathrm{dm}}\right|\beta_{p})$.
This perturbation can be neglected as long as $\left|\delta\omega_{\mathrm{dm}}\right|\!\ll\!\beta_{p}\!\sim\!0.01$.
The periapse shift due to the hypothesized particle dark matter cusp
is completely negligible (e.g. \citealt{Gne2004}). The periapse shift
in the orbit of S2 due to the observed luminous stellar mass (\citealt{Gen2003a})
is only $\left|\delta\omega_{\mathrm{dm}}\right|\!\sim\!3\!\times\!10^{-4}$.
The possible shift due to an extreme case of mass segregation of stellar
black holes (\citealt{AleLiv2004}) may be as high as $\left|\delta\omega_{\mathrm{dm}}\right|\!\sim\!10^{-3}$.

To summarize, we have shown that recent advances in infrared stellar
spectroscopy allow the decisive detection of $\beta^{2}$ post-Newtonian
effects in the Doppler shift curves with existing instruments after
about a decade of observations.

\acknowledgements{We thank Re'em Sari for important comments. TA is supported by ISF
grant 295/02-1, Minerva grant 8484, and a New Faculty grant by Sir
H. Djangoly, CBE, of London, UK.}

\end{document}